\documentclass[10pt,a4paper]{article}

\usepackage[T1]{fontenc}
\usepackage[columnsep=20pt]{geometry}
\usepackage{mathdots,physics}

\usepackage{color}
\definecolor{dark-gray}{gray}{0.20}
\definecolor{gray}{gray}{0.30}
\definecolor{light-gray}{gray}{0.80}
\definecolor{dark-red}{rgb}{0.7,0,0}
\definecolor{dark-green}{rgb}{0.1,0.4,0}
\definecolor{dark-blue}{rgb}{0.3,0.3,0.7}
\definecolor{light-blue}{rgb}{0.8,0.8,1}

\usepackage{cite}
\usepackage{hyperref}
\hypersetup{
	colorlinks=true,
	linkcolor=dark-blue,
	citecolor=dark-red,
	urlcolor=dark-green,
	linktoc=page
}
\usepackage{cuted}
\usepackage{flushend}
  \usepackage{a4wide}
  \usepackage{multicol}
  \usepackage{latexsym}
  \usepackage{epsf}
  \usepackage{bbm}
  \usepackage{graphicx}
  \usepackage{amsmath,amssymb,amsthm}
  \usepackage{verbatim}
\usepackage{comment}

\newcommand{\rme}{\mathrm e}
\newcommand{\rmd}{\mathrm d}
\newcommand{\rmi}{\mathrm i}

\newcommand{\bbm}{\left(\begin{matrix}}
\newcommand{\ebm}{\end{matrix}\right)}
\newcommand{\bea}{\begin{eqnarray}}
\newcommand{\eea}{\end{eqnarray}}
\newcommand{\be}{\begin{equation}}
\newcommand{\ee}{\end{equation}}

\numberwithin{equation}{section}
\begin{document}
\begin{center}

{\LARGE {\bf On the correspondence principle for \\\vspace{0.3cm}the Klein-Gordon and Dirac Equations}}  \\

\vspace{0.5 cm} {\large  Kevin G. Hern\'andez $^{a\,\dagger}$, Sergio E. Aguilar-Gutierrez $^{b,a\,\ddagger}$ and Jorge Bernal $^{c}$}\\
 \vspace{0.5 cm}
 
${}^a$ Department of Physics, University of El Salvador, 6224 Final, 25 Ave. Nte.,\\ Ciudad Universitaria, San Salvador, El Salvador\\
\vspace{0.3 cm}
${}^b$ Instituut voor Theoretische Fysica, K.U. Leuven,\\
Celestijnenlaan 200D, B-3001 Leuven, Belgium\\
\vspace{0.3 cm}
${}^c$ División Académica De Ciencias Básicas, Universidad Juárez Autónoma de Tabasco,\\ Carretera Cunduacán-Jalpa Km 1, Cunduacán, Tabasco, México. A.P. 24 C.P. 86690\\
\vspace{0.3 cm}
\textbf{Corresponding authors}: ${}^\dagger$kevinhernandezbel@hotmail.com; ${}^\ddagger$sergio.ernesto.aguilar@gmail.com.\\
\vspace{0.3 cm}
\textbf{Main contributions}: ${}^\dagger$Original idea. ${}^\ddagger$Elaboration of the article.
\vspace{0.5cm}
\thispagestyle{empty}
\end{center}
{\bf Abstract}

{\small  \noindent We investigate the asymptotic behavior of the solutions to the Klein-Gordon and Dirac equations using the local spatial averaging approach to Bohr's correspondence principle in the large principal quantum number regime. The procedure is applied in two basic problems in $1+1$-dimensions, the relativistic quantum oscillator and the relativistic particle in a box. In the harmonic oscillator cases, we find that the corresponding probability densities reduce to their respective classical single-particle distributions plus a series of terms suppressed by powers of the $\hbar$ constant, while particle in a box cases show a different structure for the quantum corrections.}

\vspace{0.3cm}
\noindent{\bf Keywords}

\noindent{\small Classical transition, Quantum foundations, Relativistic wave equations}

\tableofcontents
\begin{multicols}{2}
\section{Introduction}\label{sec:level1}
The recovery of classical physics from quantum observables has been one of the topics at the front of quantum foundations since its early formulation. In light of recent developments in \cite{Bernal, Martin, Martin2, Canas:2022qzs} for addressing this old problem from the local spatial, and temporal averaging (LSA and LTA respectively) procedure, our work is devoted towards calculating quantum and relativistic corrections of classical probability distributions for single particle states in terms of expansion parameters, including the Planck constant $\hbar$ and the speed of light $c$.

Perhaps the most elementary incantation of quantum to classical transition is the Planck limit, taken when $\hbar$ is negligible respect to a relevant action, which can be an useful approach to recover effective actions at leading order. A second well-established relation with the classical limit appears in the Ehrenfest theorem, which states that the quantum mechanical expectation values of the position and momentum operators  satisfy the classical equations of motion \cite{PhysRevA.66.062103, PhysRevA.70.032111}. However, the Ehrenfest theorem is neither necessary nor sufficient for a quantum system to have a {classical behavior}, since the classical limit of a quantum system is an ensemble of classical orbits where its mean position does not necessarily follow from the corresponding classical orbit \cite{PhysRevA.50.2854}. These ideas have been explored in the perspective of damped driven oscillatory systems in \cite{Choi}. Another example is Wigner's distribution function, which is an analogue of a classical probability density (PD) function over phase space that allows to study the quantum corrections of classical statistical mechanics \cite{PhysRevA.78.022109, PhysRevA.73.012104}, although it requires certain restrictions to be interpreted as a probability distribution. Lastly, a proposal with a similar vision to ours \cite{Doncheski, Berberan} compares classical and quantum single particle probability densities, although they do not establish a correspondence between such distributions.

Niels Bohr established a correspondence principle where a classical behavior is recovered when the principal quantum number, $n$, is large. Bohr applied this correspondence for frequencies and orbits of quantum systems \cite{Bernal, Liboff}, which has been applied successfully in atomic physics \cite{PhysRev.24.347,PhysRev.26.749,PhysRevA.58.186}. However, there are examples where classical frequencies can be recovered from small principal quantum numbers \cite{PhysRev.24.463}. An example of "breakdown" of the Bohr's correspondence principle in the semi-classical regime was discussed in \cite{PhysRevLett.83.4225}, which was later disputed \cite{PhysRevLett.86.2693}. Some recent studies about the emergence of the classical macroscopic realm from condensates with a large number of constituents in quantum systems can be found in \cite{Foti2019wheneverquantum, 2019FoPh49.1365H, universe7090315}.

Remarkably, one can state the correspondence principle from the LSA procedure on the quantum mechanical probabilities densities, which leads to a classical distribution for the same non-relativistic quantum systems in periodic motion \cite{robinett1995quantum,yoder2006using,rowe1987classical}. This idea has been examined for analytic examples, including the quantum harmonic oscillator, the particle in an infinite box, and the quantum Kepler problem by \cite{Bernal, Martin, Martin2}; and more recently, it has been extended to a LTA by \cite{Canas:2022qzs} for particles bouncing on the presence of a homogeneous gravitational field. However, to the best of our knowledge, the LSA and LTA procedures have only been performed in non-relativistic quantum mechanical systems.

For the first time, we adapt the LSA procedure to single particle states described by the Klein-Gordon (KG) and Dirac equations, and we apply our arguments on particular relativistic quantum systems in 1+1-dimensions, namely for the infinite square well and the harmonic oscillator. We examine the classical energy regime of our results and we find several consistency checks, the LSA PD is expressed as to leading order as the classical single-particle distribution, and {it is followed} by quantum correction terms, and we provide a discussion on the structure of such corrections for each case.

However, there are clear limitations in the LSA, which is formulated in first quantization, which means that it is limited to particles whose kinetic energy is low compared to their rest energy. For instance, it does not apply to massless particles. Thus, we do not handle non-fixed particle number effects, such as pair creation or decay, which we leave as an avenue for future research. Moreover, despite that the proposal could be extended for higher spin massive particles in the regime where first quantization is still valid, for practical reasons and for possible applications, the text is limited to the Dirac and KG equations.

Our manuscript is organized as follows. In Section \ref{sec:level2} we propose how to apply the LSA approach program in relativistic quantum mechanics, namely for the solutions to the KG and Dirac equations. In Section \ref{sec:level3} we illustrate the proposal with some simple cases where we have analytic control over the distributions in $1+1$-dimensional systems, including the KG and Dirac particles in presence of a simple harmonic oscillator and the infinite box potentials, where we recover classical probability distributions with corrections. We conclude with remarks and an outlook in Section \ref{sec:level4}.
\section{General procedure}\label{sec:level2}
\subsection{Review of local averaging procedure}\label{sec:levelReview}
To keep our article self-contained, we begin with an outline of the previous ideas involving the LSA correspondence that will carry over to the KG and Dirac equations.

The formulation of the correspondence principle in a LSA sense of the quantum PD was first realized in \cite{robinett1995quantum} (see also \cite{rowe1987classical,yoder2006using}). For clarity, we illustrate the procedure in 1+1-dimensional systems, but it can be performed in a similar way in higher dimensional non-relativistic quantum systems \cite{Martin}. Moreover, it can be extended to local time domain average instead of the spatial one in situations where there is time diffraction in matter waves \cite{Canas:2022qzs}. The classical (Cl) and quantum mechanical (QM) probability distributions, denoted by $\rho^{Cl}(x,\,t)$ and $\rho^{QM}_n(x,\,t)$ respectively, for a given system can be related in the limit of large principal quantum number $n$ by,
\begin{equation}
    \rho^{Cl}(x,\,t)=\lim_{n\rightarrow N}\frac{1}{2\epsilon_n}\int_{x-\epsilon_n}^{x+\epsilon_n}\rmd x'\,\rho_n^{QM}(x',\,t)\label{eq:rhoBohr}
\end{equation}
where $N\gg1$ is a large but finite principal quantum number defined when the energy of the QM system approaches the classical limiting value
\begin{equation}
     \lim_{n\rightarrow N} E_n^{QM}=E^{Cl}, \label{eq:Energy limit}
\end{equation}
and the parameter $\epsilon_n$ is related with the standard deviation of the particle's position, which approaches $\lim_{n\rightarrow N}\epsilon_n\rightarrow 0$.

To simplify the evaluation of the original integral (\ref{eq:rhoBohr}), \cite{Bernal, Martin, Martin2} mapped the problem to the Fourier transformed space, where instead of directly evaluating the LSA in (\ref{eq:rhoBohr}) one can evaluate the limit $n\rightarrow N$ directly in the Fourier transformed space, and then transformed back to the original space to find the LSA PD, which leads to the classical probability distribution of the system in the limit when we fix $n$ such that (\ref{eq:Energy limit}) is obeyed. This procedure proves to be convenient for periodic systems, as it has been successfully demonstrated with several examples \cite{Bernal, Martin, Martin2, Canas:2022qzs}. In these particular examples, one is able to easily interpret the asymptotic PD in terms of the classical case with a series of quantum corrections.

\subsection{Proposal for the KG and Dirac equations}
We seek to transfer the previous ideas in the context of relativistic quantum mechanics (RQM); however, it becomes more subtle to construct probability densities in this case. The antiparticle solutions of the KG equation can have negative energies and lead to negative probabilities; although there is a stochastic interpretation of quantum theory \cite{CUFAROPETRONI1984368} where both particles and antiparticles have positive energies and move forward in time, with antiparticles moving with momenta in the opposite direction to the corresponding particle, resulting in well-defined probabilities. The reader can find an useful discussion {in} \cite{nikolic2007probability} for how to construct probability densities in RQM for more general settings that what we study bellow.

In the case of $1+1$-dimensional single particle states, there is a space-like surface which {is} isomorphic to $\mathbb{R}$ or a compact interval, with a time-like conserved current $j^\mu$ given by
\begin{equation}
    j_\mu=\begin{cases}
    \rmi\,(\phi^* \,{\partial}_{\mu}\phi-\phi \,{\partial}_{\mu}\phi^*),&(\text{KG})\\
    \overline{\psi}\,\gamma_\mu\psi,& (\text{Dirac}).
    \end{cases}\label{eq:jmu}
\end{equation}
and $\phi(x,\: t)$ is the time-dependent wave function, and $j_0$ is identified {as the} relativistic PD in position space for positive energy states. Like in the non-relativistic case, we may evaluate the LSA integral from (\ref{eq:rhoBohr}) with $\rho_{n}^{QM}(x,\,t)\rightarrow\rho_{n}^{RQM}(x,\,t)$ for $n\gg1$ , and apply the asymptotic analysis in Fourier space if it were convenient, then fix the energy $E^{RQM}_n$ to its classical value (including the particle's rest energy) in (\ref{eq:Energy limit}) to recover the classical distribution $\rho_{n}^{Cl}(x,\,t)$ at leading order.

We will proceed by applying this procedure for eigenstate solutions. The KG solutions would be given by terms of the form $\phi(x,t)=\phi_n(x)\, \rme^{-\frac{\rmi E_n t}{\hbar}},$ where $\phi_n(x)$ denotes the wave function of the $n$-th eigenstate with energy $E_n$, and similarly for the Dirac equation solutions. The resulting time-independent probability densities in RQM are given by 
\begin{equation}
    \rho^{RQM}_n(x)=\begin{cases}
    \phi_n^*(x)\phi_n(x),&(\text{KG}).\\
    \psi_n^\dagger(x)\psi_n(x),&(\text{Dirac})
    \end{cases}
\end{equation}
where only the $E_n>0$ solutions are physical. The Dirac solutions automatically have a positive definite PD for single particle states, but not in the multiparticle case \cite{berndl1996nonlocality}. We illustrate this proposal in particular single particle state systems where we have analytic control.

\section{Applications}\label{sec:level3}
\subsection{Klein-Gordon oscillator}\label{Sec:KG osc}
The wave function of the 1+1-dimensional KG oscillator is exactly the same as that of the Schr\"odinger oscillator, and the PD is given by
\begin{equation}
\rho^{RQM}_n(x)=\sqrt{\frac{\alpha}{\pi}}
\frac{1}{2^n n!}[H_n(\sqrt{\alpha}x)]^2\rme^{-\alpha x^2},\label{KGO-rho1}
\end{equation}
where $H_n(x)$ are the Hermite polynomials, $n$ is a non-negative integer, and $\alpha\equiv \frac{m\omega}{\hbar}$, where $m$ is the mass of the particle or antiparticle, $\omega$ the frequency of the oscillator. The energy spectrum is given by \cite{rao},
\begin{equation}
    E_n^2=m^2c^4+2\qty(n+\frac{1}{2})mc^2\hbar \omega,
\end{equation}
where we only consider $E_n>0$ solutions. To calculate the LSA (\ref{eq:rhoBohr}), as we stated in Section \ref{sec:level1}, we transform (\ref{KGO-rho1}) to Fourier space and evaluate the asymptotic limit $n\gg1$ of the Laguerre polynomials found in the literature \cite{Grad}, after which we transform back to position space, to recover
\begin{align}
\rho^{RQM}_n(x)\sim&\frac{1}{\pi}\frac{1}{\sqrt{\kappa_n^2-x^2}}\label{KGO-rho2}\\
&+{2\pi\kappa_n}\sum_{j=1}^\infty\qty(-\frac{\pi\,\hbar^2}{32\,S_n^2})^j i_j(x,\kappa_n),\nonumber
\end{align}
where \begin{equation}
\begin{aligned}
\kappa_n&\equiv\sqrt{\frac{2\hbar(n+\frac{1}{2})}{m\omega}}=\sqrt{\frac{E_n^2-m^2c^4}{m^2\omega^2c^2}},\\
S_n&=\pi m \omega \kappa_n^2,\label{kappax0}
\end{aligned}
\end{equation}
and $i_j(x,\kappa_n)$ is the $j$-th dimensionless integral, previously found in \cite{Bernal} in the non-relativistic version of this problem. We provide a close expression for it,
\begin{align}
i_k(x,\: y)=&\int _{-\infty}^{\infty}\rmd\alpha\: \rme^{\rmi\alpha \frac{x}{y}}\int_0^\alpha \rmd\beta_1\,F(\alpha,\:\beta_1)\dots\nonumber\\
&\times\int_0^\alpha \rmd\beta_k\, F(\beta_{k-1},\:\beta_k)J_0(\beta_k)
\end{align}
where
\begin{equation}
    F(\alpha,\:\beta)\equiv\beta^3[J_0(\alpha)Y_0(\beta)-J_0(\beta)Y_0(\alpha)].
\end{equation}
Notice that the exact quantum {mechanical and the asymptotic probability densities, in (\ref{KGO-rho1}) and (\ref{KGO-rho2}) respectively,} are related through (\ref{eq:rhoBohr}) before taking the classical energy regime in (\ref{eq:Energy limit}).

Following the procedure outlined in Section \ref{sec:levelReview}, we take the classical energy limit (\ref{eq:Energy limit}) to fix $n$ and deduce the classical behavior of the system; which in the present case amounts to $E_n\rightarrow mc^2+\frac{1}{2}m\omega^2x_0^2$, where $x_0$ is the amplitude of the oscillator. In this limit with $\omega x_0\ll c$, which means $\kappa_n\rightarrow x_0$ in (\ref{KGO-rho2}). The resulting limit coincides with the non-relativistic version of the system found by \cite{Bernal}.

Eq. (\ref{KGO-rho2}) shows that the leading order result of the probability distribution is $\hbar$-independent, and it corresponds to the classical distribution of a single particle in a harmonic oscillator potential, while the higher order terms can be interpreted as quantum corrections.

\subsection{Klein-Gordon particle in a box}
Consider a KG stationary state in an infinite square well of length $0\leq x \leq L$. The PD is almost the same as the one that would be found from the Schr\"odinger equation \cite{sakurai}
\begin{equation}
\rho^{RQM}_n(x)=\frac{2}{L}\sin^2\qty(\frac{n\pi x}{L}).\label{KGBox exact}
\end{equation}
The difference from the non-relativistic case is the energy spectrum, which is given by
\begin{eqnarray}
E_n^2=m^2c^4+c^2\hbar^2\frac{n^2\pi^2}{L^2}.
\end{eqnarray}
We computed the LSA of (\ref{KGBox exact}) as in the previous example, resulting in
\begin{equation}
\rho^{RQM}_n(x)\sim\frac{1}{L}\qty[H(L-x)-H(-x)],\label{PD-ISW}
\end{equation}
where $H(x)$ is the Heaviside step function. We observe that the KG {PD is the same as the one in the non-relativistic version of this system \cite{Bernal}.} There are no quantum correction terms appearing on the distribution, and the energies do not need to be fixed to a classical value. This result will be contrasted with the corresponding Dirac particle problem, where quantum corrections are manifest.

\subsection{Dirac oscillator}
The Dirac oscillator was originally proposed by Moshinsky and Szczepaniak \cite{Moshinsky}, which is an exactly solvable model of Dirac fermions subject to a Hamiltonian whose square is the same as the Hamiltonian for KG particles with a harmonic oscillator interaction, plus a spin-orbit interaction that does not appear in the $1+1$-dimensional case. The eigenfunctions in the 1-dimensional case can be found in \cite{Radoslaw}. For either particles or antiparticles, the PD for single eigenstates (either spin up or down) has the form
\begin{equation}
\begin{aligned}
\rho^{RQM}_n(x)=&\rme^{-\alpha x^2}
\abs{a_{n}}^2 H_n^2(\sqrt{\alpha}x)\\
&+\rme^{-\alpha x^2}\abs{a'_{n}}^2 H_{n-1}^2(\sqrt{\alpha}x),
\end{aligned}
\label{DO-rho1}
\end{equation}
where $\alpha\equiv\frac{m\omega}{\hbar}$, $E_n^2=m^2c^4+2n\hbar \omega mc^2$ and
\begin{align}
    \abs{a_{n}}^2&=\frac{\sqrt{\alpha}(E_n+mc^2)}{2^{n+1} n!E_n\sqrt{\pi}},\\ \abs{a'_{n}}^2&=\frac{\sqrt{\alpha}(E_n-mc^2)}{2^{n} (n-1)!E_n\sqrt{\pi}}.
\end{align}
One can evaluate the LSA of Eq. (\ref{DO-rho1}),
\begin{align}
&\rho^{RQM}_n(x)\sim\label{DO-rho2}\\ &\frac{1}{\pi}\abs{\frac{a'_{n}}{A_{n-1}}}^2\frac{1}{\sqrt{\kappa_{n-1}^2-x^2}}+\frac{1}{\pi}\abs{\frac{a_{n}}{A_{n}}}^2\frac{1}{\sqrt{\kappa_{n}^2-x^2}}\nonumber\\
&+\abs{\frac{a'_{n}}{A_{n-1}}}^2\frac{1}{2\pi\kappa_{n-1}}\sum_{j=1}^\infty\qty(-\frac{\pi\,\hbar^2}{32\,S_{n-1}^2})^j i_j(x,\kappa_{n-1})\nonumber\\
&+\abs{\frac{a_{n}}{A_{n}}}^2\frac{1}{2\pi\kappa_{n}}\sum_{j=1}^\infty\qty(-\frac{\pi\,\hbar^2}{32\,S_{n}^2})^j i_j(x,\kappa_{n})\,,\nonumber
\end{align}
where $\abs{A_{n}}^2=\sqrt{\frac{\alpha}{\pi}}\frac{1}{2^n n!}$
, while $\kappa_n$ and $S_n$ are shown in (\ref{kappax0}). There is a particular feature of the Moshinsky model for the Dirac oscillator that we need to consider before taking the non-relativistic limit. The Moshinsky model does not reproduce the non-relativistic energy values that would be found in quantum harmonic oscillator from the Schr\"odinger equation, because of how the harmonic term is added in the Hamiltonian \cite{Moshinsky}, such that in the non-relativistic limit, the energy spectrum contains a factor $n\hbar\omega$ instead of $(n+\frac{1}{2})\hbar \omega$ \cite{Radoslaw}. This means that the solutions in this model have $n$ half integer. In order to consider only integer quantum numbers, we can work with a shifted variable $N=n\pm\frac{1}{2}$. In this way the non-relativistic energy spectrum depends on an integer quantum number, as in the Schr\"odinger case. {This modification is only a matter of convenience.}

We {fix the energy} $E_{N}\rightarrow mc^2+\frac{1}{2}m\omega^2x_0^2$ to make connection with the classical limit, where we have $\abs{a_{n}}^2 \rightarrow \abs{A_n}^2$, $a'_n\rightarrow 0$ and $\kappa_{n}\rightarrow x_0$. In that case, (\ref{DO-rho2}) becomes (\ref{KGO-rho2}) with $\kappa_n\rightarrow x_0$, which agrees with the non-relativistic version of the problem studied by \cite{Bernal}. {The result is independent on whether the state is in a state with spin up or down, or a linear combination of both.}

{We conclude this part of the section with the interpretation of our results so far.} The classical limit for the probability distribution of Dirac fermions is well-known in the case of condensates, i.e. transition from Fermi-Dirac distribution to Maxwell-Boltzmann in case of high temperatures and low particle density. However, our notion of the classical limit at the level of probability distribution is done for single particle states at high enough energies to reach the classical regime in (\ref{eq:Energy limit}).{ Therefore, fine-grained details of the system such as its intrinsic spin are coarse-grained away in the LSA approach to the correspondence principle, where the degree of the coarse-graining will be reflected in the quantum correction series remaining from (\ref{DO-rho2}).}

\subsection{Dirac particle in a box}
Let us consider the Dirac particle wave function,  $\psi^{(+)}_k(x)$, for a 1-dimensional infinite square Lorentz  scalar  potential found explicitly in \cite{alb}, where there is no Klein paradox \footnote{In the Klein paradox, a strong vector potential leads to a non-zero transmission to a classically forbidden region, which is related to particle/anti-particle pair production. To avoid it, \cite{alb} considered a scalar potential defined in a one-dimensional box, and since it is invariant under Lorentz transformations, this issue doesn't arise in such case.}. The corresponding antiparticle solution, $\psi^{(-)}_k(x)$, can be calculated as  $\psi^{(-)}_k(x)=\gamma^5\psi^{(+)}_{-k}(-x)$, where $\gamma^5=\rmi\gamma^0\gamma^1\gamma^2\gamma^3$. {The PD for} either particle or antiparticle, with spin up or down, has form
\begin{align}
\rho^{RQM}_k(x)=&\abs{B_k}^2 \cos^2\qty(kx-\frac{\delta_k}{2})\label{eq:rhoDPIB}\\
&+\abs{B_k}^2 \Phi_k^2\sin^2\qty(kx-\frac{\delta_k}{2}),\nonumber
\end{align}
for the interval $0\leq x \leq L$, and the PD vanishes outside this interval. We have also introduced the following definitions
\begin{align}
&\abs{B_k}^2\equiv \tfrac{4k}{(\Phi_k^2-1)(2kL-\sin(kL+\delta_k)-\sin\delta_k)+4k\,L},\nonumber\\
&k\equiv \frac{1}{\hbar}\sqrt{\frac{E_k^2}{c^2}-m^2c^2},\quad\Phi_k\equiv \frac{\hbar k c}{E_k+mc^2},\nonumber\\
&\delta_k\equiv \arctan\qty(\frac{2\Phi_k}{\Phi_k^2-1}).
\end{align}
The energy of the particle $E_k$ can be found from the condition $\tan(kL)=-\frac{\hbar k}{mc^2}$ \cite{alb}. We evaluate the asymptotic limit, i.e. $kL\gg 1$, to find the LSA of (\ref{eq:rhoDPIB}) as
\begin{equation}
\rho^{RQM}_k(x)\sim\frac{1+\Phi_k^2}{2}\abs{B_k}^2\qty[H(L-x)-H(-x)].
\end{equation}
It should be noticed that the terms $\Phi_k^2$ and $\abs{B_k}^2$ of the PD appear as fermionic quantum parameters, given that Eq. (\ref{PD-ISW}) for the Klein Gordon system does not contain similar factors.

The non-relativistic limit can be stated in the condition $\hbar k\ll mc$, which implies that $\Phi_k\rightarrow 0$, $\abs{B_k}^2\rightarrow 2/L$, and the PD becomes (\ref{PD-ISW}). The result is similar to the corresponding KG solution in the sense that the non-relativistic LSA PD does not include a series of quantum corrections, because there is no dependence on $\hbar$.

As a remark, there is a similar type of Dirac particle in a box model in \cite{alhaidari2009dirac} where the potential is the time component of a vector, and the considerations of the Klein paradox have been studied in detail. We expect that our procedure in that case would probably lead to similar results as what we found here.

\section{Conclusions}\label{sec:level4}
To summarize, we reformulated the LSA approach to the correspondence principle for the KG and Dirac equations. {To the best of our knowledge, this approach had only been studied so far in non-relativistic quantum systems.} In both cases, we obtained classical PD with quantum corrections for the infinite well and the quantum harmonic oscillator potentials.

For the solutions of the KG equation, the PD coincides with the ones derived from the Schrödinger equation for the corresponding potential term \cite{Bernal}, although the energy spectrum is modified. {In contrast, both the PD and energy spectrum of the Dirac equation solutions differ from those studied in \cite{Bernal}. Moreover,} we interpreted the macroscopic limit of single particle Dirac fermions from the point of view of its PD at the high energy regime; in a similar way to what occurs for condensates transitioning from a Fermi-Dirac distribution to a Maxwell-Boltzmann distribution.

For both the KG and the Dirac oscillators, we found that the quantum corrections to the classical PD were expressed as power series in the $\hbar$ constant, which we verified to be negligible respect to the classical contributions. The LSA PD for a particle in a box potential showed a different structure for the quantum corrections. In the KG case showed no corrections, while the Dirac fermion solution had an amplitude modulation that depends on $\hbar$ and $c$ constants.

Let us discuss possible future research directions. Motivated by the recent study in \cite{Canas:2022qzs} where the LTA procedure is applied to propose tests of the equivalence principle, it might be interesting to study other experimentally accessible systems involving the classical transition with (special) relativistic corrections. A first step in this direction is to apply our proposal to the relativistic hydrogen atom problem, which we are currently exploring. On related direction, there are instances when electrons might travel at high speeds but their measured speed is much lower due to run-and-tumble stochastic effects \cite{Maes:2021tkd} and they can be described with first quantization, which might provide a testing ground of the LSA procedure for relativistic particles; for instance, to describe the trajectory of the electron's trajectory in single and double slit experiments as the macroscopic limit is approached.

\section*{Acknowledgements}
We thank Kwinten Frazen and Christian Maes for their detailed comments in the draft version of our manuscript; and to Mauricio Paulin and Alberto Ruiz for useful discussions and suggestions. The work of S.E.A.G. is supported by the KU Leuven grant C16/16/005. Furthermore, S.E.A.G thanks IFT-UNESP \& ICTP-Trieste where part of this work was performed.

\section*{Conflicts of interest}
The authors declare no conflicts of interest.

\section*{Abbreviations}{
The following abbreviations are used in this manuscript:\\

\noindent 
\begin{tabular}{@{}ll}
{PD} & {Probability density/ distribution}\\
LSA & Local spatial averaging/ averaged\\
LTA & Local temporal averaging/ averaged\\
KG & Klein-Gordon\\
Cl & Classical mechanical\\
QM & Quantum mechanical\\
RQM & Relativistic quantum mechanics
\end{tabular}
}
\setcounter{tocdepth}{2}

\providecommand{\href}[2]{#2}\raggedright
\end{multicols}
\end{document}